\documentclass[onecolumn,12pt,draft,a4paper]{article}
\usepackage{amsmath}
\usepackage{aip}
\usepackage{a4}
\usepackage{graphicx}
\usepackage{amsfonts}
\usepackage{amssymb}

\begin{document}

\author{Dingping. Li\\\textit{National Center for Theoretical Sciences} \\\textit{P.O.Box 2-131, Hsinchu, Taiwan, R. O. C.}
\and Baruch Rosenstein\\\textit{National Center for Theoretical Sciences} \\\textit{\ and} \textit{Electrophysics Department}, \\\textit{National Chiao Tung University } \\\textit{Hsinchu 30043, Taiwan, R. O. C.}}
\title{Why the lowest Landau level approximation works in strongly type II superconductors.}
\date{\today}
\maketitle
\begin{abstract}
Higher than the lowest Landau level contributions to magnetization and
specific heat of superconductors are calculated using Ginzburg - Landau
equations approach. Corrections to the excitation spectrum around solution of
these equations (treated perturbatively) are found. Due to symmetries of the
problem leading to numerous cancellations the range of validity of the LLL
approximation in mean field is much wider then a naive range and extends all
the way down to $H=\frac{H_{c2}(T)}{13}$. Moreover the contribution of higher
Landau levels is significantly smaller compared to LLL than expected naively.
We show that like the LLL part the lattice excitation spectrum at small
quasimomenta is softer than that of usual acoustic phonons. This enhanses the
effect of fluctuations. The mean field calculation extends to third order,
while the fluctuation contribution due to HLL is to one loop. This complements
the earlier calculation of the LLL part to two loop order.
\end{abstract}

\section{Introduction}

Ginzburg - Landau effective description of high $T_{c}$ superconductors has
been remarkably successful in describing various thermodynamical and transport
properties \cite{Blatter}. However when fluctuations are of importance, even
this effective description becomes very complicated. Some progress can be
achieved when certain additional assumptions are made. One of the often made
additional assumption is that only the lowest Landau level (LLL) significantly
contributes to physical quantities of interest
\cite{Thouless,Eilenberger,Moore1,Moore2,Tesanovic1,Sasik,Rosenstein}. There
is a debate however on how restrictive the LLL approximation actually is.
Naively when $H<\frac{H_{c2}(T)}{3}$ (see the dotted line on Fig. 1), even
within mean field approximation, one should consider higher Landau levels
(HLL) mixing in the Abrikosov vortex lattice solution of the GL equations.
When fluctuations are included one can argue using Hartree approximation
\cite{Lawrie} that the LLL range of validity is even smaller. However direct
application of the LLL scaling to magnetization and specific heat on YBCO
suggest that the range of applicability is much wider - all the way down to
$1-3T$ \cite{Pierson,Sasik,Tesanovic2}. It is not clear why HLL do not contribute.

In this paper we explicitly calculate the effects of HLL at low temperatures
in the vortex solid or liquid phase and establish the realistic range where
the LLL approximation is valid (see the heavy dashed line on Fig.1). We
reanalyze the HLL corrections to mean field equations going to higher order
then in \cite{Lascher} and find that the expansion converges for
$H_{c2}>H>\frac{H_{c2}}{13}$. Importantly within this radius of convergence
the LLL contribution constitutes more than $95\%$. Then we calculate the HLL
fluctuation effects to one loop order complementing the LLL calculation to two
loops by one of us \cite{Rosenstein} (later referred to as I).

Ginzburg parameter $Gi$ characterizing importance of thermal fluctuations is
much larger in high $T_{c}$ superconductors then in the low temperature ones.
Moreover in the presence of magnetic field the importance of fluctuations in
high $T_{c}$ superconductors is further enhanced. Under these circumstances
corrections to various physical quantities like magnetization or specific heat
are not negligible even at low temperatures. It is quite straightforward to
systematically account for the fluctuations effect on magnetization, specific
heat or conductivity perturbatively above the mean field transition line using
Ginzburg - Landau description \cite{Tinkham}. However in the interesting
region below this line it turned out to be extremely difficult to develop a
quantitative theory.

Within LLL in order to approach the region below the mean field transition
line $T<T_{mf}(H)$ Thouless \cite{Thouless} proposed a perturbative approach
around homogeneous (liquid) state was in which all the ''bubble'' diagrams are
resumed. The series provide accurate results at high temperatures, but for the
LLL dimensionless temperature $a_{T}\equiv\left(  \frac{2H_{c2}^{2}}%
{GiT_{c}T^{2}H^{2}}\right)  ^{1/3}\frac{T-T_{mf}(H)}{\pi}\lesssim-2$ become
inapplicable. Generally attempts to extend the theory to lower temperature by
Pade extrapolation were not successful \cite{Moore1}. Alternative, more direct
approach to low temperature fluctuations physics is to start from the mean
field solution and then take into account perturbatively fluctuations around
this inhomogeneous solution. Experimentally it is reasonable since, for
example, specific heat at low temperatures is a smooth function and the
fluctuations contribution experimentally is quite small. For some time this
was in disagreement with theoretical expectations.

Eilenberger calculated spectrum of harmonic excitations of the triangular
vortex lattice (see eq.(\ref{elspec}) below) \cite{Eilenberger} and noted that
the gapless mode is softer then the usual Goldstone mode expected as a result
of spontaneous breaking of translational invariance. The inverse propagator
for the ''phase'' excitations behaves as $k_{z}^{2}+const(k_{x}^{4}+k_{y}%
^{4})$. The influence of this unexpected additional ''softness'' apparently
goes beyond enhancement of the contribution of fluctuations at leading order.
It leads to disastrous infrared divergencies at higher orders rendering the
perturbation theory around the vortex state doubtful. One therefore tends to
think that nonperturbative effects are so important that such a perturbation
theory should be abandoned \cite{Ruggeri}. However it was shown in I that a
closer look at the diagrams reveals that in fact one encounters actually only
logarithmic divergencies. This makes the divergencies similar to so called
''spurious'' divergencies in the theory of critical phenomena with broken
continuous symmetry and they exactly cancel at each order provided we are
calculating a symmetric quantity. Qualitatively physics of fluctuating $D=3$
GL model in magnetic field turns out to be similar to that of spin systems in
$D=2$ possessing a continuous symmetry. In particular, although within
perturbation theory in thermodynamic limit the ordered phase (solid) exists
only at $T=0$, at low temperatures liquid differs very little in most aspects
from solid. One can effectively use properly modified perturbation theory to
quantitatively study various properties of the vortex liquid phase. This
perturbative approach agrees very well with the direct Monte Carlo simulation
of \cite{Sasik}. The question arises whether one can extend the well
controlled perturbative calculation beyond the LLL. Sometimes a hope is
expressed that the additional softness is an accidental artifact of LLL
approximation. The present work explicitly shows that this is not so. It is a
fundamental general phenomenon.

The paper is organized as follows. Model is described and perturbative mean
field solution developed in section II. The expansion parameter will be the
distance from the mean field critical line $a_{h}\equiv\frac{1}{2}(1-\frac
{T}{T_{c}}-\frac{H}{H_{c2}})$. Range of validity of the expansion and of the
LLL approximation is discussed. Then in section III we derive the spectrum of
excitations to leading order and to the next to leading order in $a_{h}$. The
free energy to one loop is calculated in section IV. Section V contains
expressions for magnetization and specific heat and discussion of validity
range of the fluctuation contributions calculation. Finally we summarize the
results in section VI. Details of the mean field calculation can be found in
Appendix A, while details of the HLL spectrum calculation can be found in
Appendix B.

\section{Model and the perturbative mean field solution}

\subsection{Model\bigskip}

Our starting point is the GL free energy:
\begin{equation}
F=\int d^{3}x\frac{{\hbar}^{2}}{2m_{ab}}|(\vec{\nabla}-\frac{ie^{\ast}}{\hbar
c}\vec{A})\psi|^{2}+\frac{{\hbar}^{2}}{2m_{c}}|\partial_{z}\psi|^{2}%
+a|\psi|^{2}+\frac{b^{\prime}}{2}|\psi|^{4} \label{energy}%
\end{equation}
Here $\vec{A}=(-By,0)$ describes a nonfluctuating constant magnetic field. For
strongly type II superconductors ($\kappa\sim100$) far from $H_{c1}$(this is
the range of interest in this paper) magnetic field is homogeneous to a high
degree due to superposition from many vortices. For simplicity we assume
$a=\alpha(1-t)$, $t\equiv T/T_{c}$ although this dependence can be easily
modified to better describe the experimental coherence length.

Throughout most of the paper will use the following units. Unit of length is
$\xi=\sqrt{{\hbar}^{2}/\left(  2m_{ab}\alpha T_{c}\right)  }$ and unit of
magnetic field is $H_{c2}$, so that dimensionless magnetic field is $b\equiv
B/H_{c2}$. The dimensionless Boltzmann factor in these units is (the order
parameter field is rescaled as $\psi^{2}\rightarrow\frac{2\alpha T_{c}%
}{b^{\prime}}\psi^{2}$):%

\begin{equation}
\frac{F}{T}=\frac{1}{\omega}\int d^{3}x\frac{1}{2}|D\psi|^{2}+\frac{1}%
{2}|\partial_{z}\psi|^{2}-\frac{1-t}{2}|\psi|^{2}+\frac{1}{2}|\psi|^{4},
\label{energ1}%
\end{equation}
The dimensionless coefficient is
\begin{equation}
\omega=\sqrt{2Gi}\pi^{2}t. \label{omega}%
\end{equation}
where the Ginzburg number is defined by $Gi\equiv\frac{1}{2}(\frac{32\pi
e^{2}\kappa^{2}\xi T_{c}\gamma^{1/2}}{c^{2}h^{2}})^{2}$ and $\gamma\equiv
m_{c}/m_{ab}$ is an anisotropy parameter. This coefficient determines the
strength of fluctuations, but is irrelevant as far as mean field solutions are concerned.

\subsection{\bigskip Mean field solution by expansion in $a_{h}$}

Now we turn to a perturbative solution of the Ginzburg-Landau equations near
the mixed state - normal phase transition line. This has been done before
\cite{Lascher} to the second order, however the range of applicability and
precision of the LLL approximation at large $\kappa$ has not been fully
explored. The $z$ direction dependence of the solutions is trivial and will
not be mentioned until fluctuations will be discussed. The expansion parameter
is
\begin{equation}
a_{h}\equiv\frac{1-t-b}{2}. \label{ah}%
\end{equation}
Rewriting the quadratic part in terms of operator (''Hamiltonian'' )
$\mathcal{H}\equiv$ $\frac{1}{2}(-D^{2}-b)$ whose spectrum starts from zero,
one obtains the following free energy density over $T$
\begin{equation}
\frac{F}{T}\equiv\frac{f}{\omega}=\frac{1}{\omega}\int d^{2}x\left(
\psi^{\ast}\mathcal{H}\psi-a_{h}|\psi|^{2}+\frac{1}{2}|\psi|^{4}\right)  .
\label{den}%
\end{equation}
The equation of motion is therefore%

\begin{equation}
\mathcal{H}\psi-a_{h}\psi+\psi|\psi|^{2}=0 \label{eqmt}%
\end{equation}
This equation is solved perturbatively in $a_{h}$ by assuming%

\begin{equation}
\Phi=(a_{h})^{1/2}\left[  \Phi_{0}+a_{h}\Phi_{1}+...\right]  \label{phi}%
\end{equation}
It is convenient to represent $\Phi_{0},\Phi_{1},...$ in the basis of
eigenfunctions of $\mathcal{H}$, $\mathcal{H}\varphi^{n}=nb\varphi^{n}$,
normalized to unit ''Cooper pairs density'' $<|\varphi^{n}|^{2}>\equiv
\int_{cell}d^{2}x|\varphi^{n}|^{2}\frac{b}{2\pi}=1$, where ''cell'' is a
primitive \ \ \ cell of the vortex lattice. Assuming hexagonal lattice
symmetry one explicitly has;%

\begin{align*}
\varphi^{n}  &  =\sqrt{\frac{2\pi}{\sqrt{\pi}2^{n}n!a}}\sum\limits_{l=-\infty
}^{\infty}H_{n}(y\sqrt{b}-\frac{2\pi}{a}l)\\
&  \exp\left\{  i\left[  \frac{\pi l(l-1)}{2}+\frac{2\pi\sqrt{b}}{a}lx\right]
-\frac{1}{2}(y\sqrt{b}-\frac{2\pi}{a}l)^{2}\right\}
\end{align*}
where $\frac{a}{\sqrt{b}}=\sqrt{\frac{4\pi}{\sqrt{3}b}}$ is the lattice spacing.

To order zero,
\begin{equation}
\mathcal{H}\Phi_{0}=0
\end{equation}
and $\Phi_{0}$ is proportional to the Abrikosov vortex lattice solution
$\varphi$ which is the expression eq.(\ref{eqmt}) for $n=0$:%
\[
\Phi_{0}=g_{0}\varphi.
\]
To order $k$, one expands
\begin{equation}
\Phi_{i}=g_{i}\varphi+\sum_{n=1}^{\infty}g_{i}^{n}\varphi^{n}.
\end{equation}
Inserting into eq.(\ref{eqmt}) one obtains to order $a_{h}^{3}$:%

\begin{equation}
\mathcal{H}\Phi_{1}=g_{0}\varphi-g_{0}|g_{0}|^{2}\varphi|\varphi|^{2}%
\end{equation}
Taking the inner product with $\varphi$ one finds that%

\begin{equation}
g_{0}=\frac{1}{\sqrt{\beta_{A}}},
\end{equation}
where the Abrikosov's constant is the following average over primitive cell:
$\beta=\beta_{A}\equiv<|\varphi|^{4}>\approx1.16$ .\ Inner product with
$\varphi^{n}$ determines $g_{1}^{n}$:
\begin{equation}
g_{1,n}=-\frac{\beta^{n}}{nb\beta^{3/2}}, \label{gn}%
\end{equation}
where $\beta^{n}\equiv<|\varphi|^{2}\varphi^{n}\varphi^{\ast}>.$ To find
$g_{1}$ we need in addition also the order $a_{h}^{5/2}$ equation:
\begin{equation}
\mathcal{H}\Phi_{2}=\Phi_{1}-(g_{0})^{2}(2\Phi_{1}|\varphi|^{2}+\Phi_{1}%
^{\ast}\varphi^{2}) \label{eq2}%
\end{equation}
Inner product with $\varphi$ gives:
\begin{equation}
g_{1}=\frac{3}{2}\sum_{n=1}^{\infty}\frac{(\beta^{n})^{2}}{nb\beta^{5/2}}.
\label{g1}%
\end{equation}

\subsection{Mean field result for free energy. Orders $a_{h}^{2}$ and
$a_{h}^{3}$}

The mean field expression for the free energy to order $a_{h}^{2}$ is well
known. Inserting the next correction eq.(\ref{phi}) into eq.(\ref{den}) one
obtains the free energy density:%

\begin{equation}
\frac{\mathcal{F}_{mf}}{T}=\frac{1}{\omega}\left[  -\frac{a_{h}^{2}}{2\beta
}-\frac{a_{h}^{3}}{\beta^{3}b}\sum_{n=1}^{\infty}\frac{(\beta^{n}){}^{2}}%
{n}\right]  =\frac{1}{\omega}\left[  -.43a_{h}^{2}-.0078\frac{a_{h}^{3}}%
{b}\right]  \label{fmf}%
\end{equation}

It is interesting to note that $\beta_{n}\neq0$ only when $n=6j$, where $j$ is
an integer. This is due to hexagonal symmetry of the vortex
lattice\cite{Lascher}. For $n=6j$ it decreases very fast with $j$: $\beta
_{6}=-.2787,\beta_{12}=.0249.$ Because of this the coefficient of the next to
leading order is very small (additional factor of $6$ in the denominator). We
might preliminarily conclude therefore that the perturbation theory in $a_{h}$
works much better that might be naively anticipated (see dashed line on Fig.1)
and can be used very far from transition line. If we demand that the
correction is smaller then the main contribution the corresponding line on the
phase diagram will be $b=.015\cdot(1-t)$. For example the LLL melting line
corresponds to $a_{h}\sim1$. This overly optimistic conclusion is however
incorrect as calculation of the following term in Appendix A shows.

\subsection{\bigskip Range of applicability of the expansion. How precise is LLL?}

Now we discuss in what region of the parameter space the expansion outlined
above can be applied. First of all note that all the contributions to
$\Phi_{1}$ are proportional to $\frac{1}{b}$. This is a general feature: the
actual expansion parameter is $\frac{a_{h}}{b}$. One can as whether the
expansion is convergent and, if yes, what is its radius of convergence.
Looking just at the leading correction and comparing it to the LLL one gets a
very optimistic estimate. For this purpose we calculated higher orders
coefficients in Appendix A. The results for the $\Phi_{2}$ are following:
\begin{equation}
g_{2}^{n}=\frac{1}{nb}\left[  g_{1}^{n}-\frac{1}{\beta}\sum_{i=0}^{\infty
}g_{1}^{i}(2<n,0|i,0>+<0,0|i,n>)\right]  \label{g2n}%
\end{equation}

and%

\begin{equation}
g_{2}=-\frac{3}{2\beta}\beta^{n}g_{2}^{n}-\frac{1}{2\sqrt{\beta}}\sum
_{i,j=0}^{\infty}g_{1}^{i}g_{1}^{j}(<0,0|i,j>+2<j,0|i,0>) \label{g2}%
\end{equation}
where $<i_{1},i_{2}|j_{1},j_{2}>\equiv<\varphi^{\ast i_{1}}\varphi^{\ast
i_{2}}\varphi^{j_{1}}\varphi^{j_{2}}>$ and $g_{j}^{i}$ when $i=0$ is defined
to be equal to $g_{j,}\varphi^{i}=\varphi$ when $i=0.$

We already can see that $g_{2}^{n}$ and $g_{2}$ are proportional to $g_{1}%
^{n}$ and in addition there is a factor of $1/n$. Since, due to hexagonal
lattice symmetry all the $g_{1}^{n}$ ,$n\neq6j$ vanish, so do $g_{2}^{n}$. We
checked that there is no more small parameters, so we conclude that the
leading order coefficient is much larger than first (factor $6\cdot5$), but
the second is only $6$ times larger than the third.

The correction to free energy is%
\begin{equation}
\frac{\mathcal{F}_{mf}}{T}=\frac{1}{\omega}\cdot\frac{0.056}{6^{2}}\frac
{a_{h}^{4}}{b^{2}} \label{en4}%
\end{equation}
Accidental smallness by factor $1/6$ of the coefficients in the $\frac{a_{h}%
}{b}$ expansion due to symmetry means that the range of validity of this
expansion is roughly $a_{h}<6b$ or $H<\frac{H_{c2}}{13}$. Moreover additional
smallness of all the HLL corrections compared to the LLL means that they
constitute just several percent of the correct result inside the region of
applicability. To illustrate this point we plot on the Fig.2 the
perturbatively calculated solution for $h=.1,t=.5$. One can see that although
the leading LLL function has very thick vortices (Fig. 2a), the first nonzero
correction makes them of order of the coherence length (Fig. 2b). Following
correction of the order $\left(  \frac{a_{h}}{b}\right)  ^{2}$ makes it
practically indistinguishable from the numerical solution. Amazingly the order
parameter between the vortices approaches its vacuum value. Paradoxically
starting from the region close to $H_{c2}$ the perturbation theory knows to
correct the order parameter so that it looks very similar to the London
approximation (valid only close to $H_{c1}$) result \ of well separated vortices.

We conclude therefore that the expansion in $a_{h}/b$ works in the mean field
better that one can naively expect. In the next section we investigate whether
the same is true for the fluctuation contribution.

\section{\bigskip Fluctuations spectrum}

\subsection{Fluctuations to leading order in $a$}

To find an excitation spectrum in harmonic approximation one expands free
energy functional around the solution found in the previous section. Within
the LLL approximation this has been done by \cite{Eilenberger}. We generalize
it to the case of all the Landau levels when perturbations due to nonlinear
term are included. The fluctuating order parameter field $\psi$ should be
divided into a nonfluctuating (mean field) part and a small fluctuation
\begin{equation}
\psi(x)=\Phi(x)+\chi(x).
\end{equation}
The energy eq.(\ref{den}) is then expanded in $\chi$ retaining only quadratic terms%

\begin{equation}
f_{2}\equiv\int d^{2}x\left[  \chi^{\ast}\mathcal{H}\chi-a_{h}|\chi
|^{2}+2|\Phi|^{2}|\chi|^{2}+\frac{1}{2}(\Phi^{\ast2}\chi^{2}+\Phi^{2}%
\chi^{\ast2})\right]  \label{f2}%
\end{equation}
Field $\chi$ can be expanded in a basis of quasimomentum $\overrightarrow{k}$eigenfunctions:%

\begin{align*}
\varphi_{\vec{k}}^{n}  &  =\sqrt{\frac{2\pi}{\sqrt{\pi}2^{n}n!a}}%
\sum\limits_{l=-\infty}^{\infty}H_{n}(y\sqrt{b}+k_{x}-\frac{2\pi}{a}l)\\
&  \exp\left\{  i\left[  \frac{\pi l(l-1)}{2}+\frac{2\pi(\sqrt{b}x-k_{y})}%
{a}l-\sqrt{b}xk_{x}\right]  -\frac{1}{2}(y\sqrt{b}+k_{x}-\frac{2\pi}{a}%
l)^{2}\right\}
\end{align*}
In addition instead of complex field $\chi_{k}^{n}$we will use two ''real''
fields $O_{k}^{n}$ and $A_{k}^{n}$ satisfying $O_{k}^{n}=O_{-k}^{\ast n}%
$,$A_{k}^{n}=A_{-k}^{\ast n}:$%

\begin{align}
\chi(x)  &  =\int_{\vec{k}.}\sum_{n=0}^{\infty}\varphi_{\overrightarrow{k}%
}^{n}(x)\left(  O_{k}^{n}+iA_{k}^{n}\right) \label{transf}\\
\chi^{\ast}(x)  &  =\int_{\vec{k}.}\sum_{n=0}^{\infty}\varphi_{\overrightarrow
{k}}^{\ast n}(x)\left(  O_{-k}^{n}-iA_{-k}^{n}\right) \nonumber
\end{align}

In terms of these fields representing ''optical'' and ''acoustic'' phonons
eq.(\ref{f2}) takes a form:%

\begin{align}
f_{2}  &  =\int_{\vec{k}.}\sum_{n=1}^{\infty}2(nh-a_{h})(O_{k}^{n}O_{-k}%
^{n}+A_{k}^{n}A_{-k}^{n})-2a_{h}(O_{k}O_{-k}+A_{k}A_{-k})\\
&  +2\sum_{i,j=0}^{\infty}A_{k}^{i}A_{-k}^{j}K_{k}^{i,j}+O_{k}^{i}A_{-k}%
^{j}L_{k}^{i,j}+O_{-k}^{i}A_{k}^{j}M_{k}^{i,j}+O_{k}^{i}O_{-k}^{j}N_{k}%
^{i,j}\nonumber
\end{align}

where elements of the matrix%

\begin{align}
K_{k}^{i,j}  &  =<|\Phi|^{2}(\varphi_{-\overrightarrow{k}}^{\ast i}%
\varphi_{-\overrightarrow{k}}^{j}+\varphi_{\overrightarrow{k}}^{i}%
\varphi_{\overrightarrow{k}}^{\ast j})-\frac{1}{2}(\Phi^{\ast2}\varphi
_{\overrightarrow{k}}^{i}\varphi_{-\overrightarrow{k}}^{j}+\Phi^{2}%
\varphi_{-\overrightarrow{k}}^{\ast i}\varphi_{\overrightarrow{k}}^{\ast
j})>\nonumber\\
N_{k}^{i,j}  &  =<|\Phi|^{2}(\varphi_{-\overrightarrow{k}}^{\ast i}%
\varphi_{-\overrightarrow{k}}^{j}+\varphi_{\overrightarrow{k}}^{i}%
\varphi_{\overrightarrow{k}}^{\ast j})+\frac{1}{2}(\Phi^{\ast2}\varphi
_{\overrightarrow{k}}^{i}\varphi_{-\overrightarrow{k}}^{j}+\Phi^{2}%
\varphi_{-\overrightarrow{k}}^{\ast i}\varphi_{\overrightarrow{k}}^{\ast
j})>\label{k}\\
L_{k}^{i,j}  &  =<|\Phi|^{2}(\varphi_{-\overrightarrow{k}}^{\ast i}%
\varphi_{-\overrightarrow{k}}^{j}-\varphi_{\overrightarrow{k}}^{i}%
\varphi_{\overrightarrow{k}}^{\ast j})+\frac{1}{2}(\Phi^{\ast2}\varphi
_{\overrightarrow{k}}^{i}\varphi_{-\overrightarrow{k}}^{j}-\Phi^{2}%
\varphi_{-\overrightarrow{k}}^{\ast i}\varphi_{\overrightarrow{k}}^{\ast
j})>\nonumber\\
M_{k}^{i,j}  &  =<|\Phi|^{2}(\varphi_{-\overrightarrow{k}}^{\ast i}%
\varphi_{-\overrightarrow{k}}^{j}-\varphi_{\overrightarrow{k}}^{i}%
\varphi_{\overrightarrow{k}}^{\ast j})+\frac{1}{2}(\Phi^{\ast2}\varphi
_{\overrightarrow{k}}^{i}\varphi_{-\overrightarrow{k}}^{j}-\Phi^{2}%
\varphi_{-\overrightarrow{k}}^{\ast i}\varphi_{\overrightarrow{k}}^{\ast
j})>\nonumber
\end{align}

\bigskip We expand $f_{2}$ in $a_{h}$. The order $a_{h}$ term is%

\begin{align}
&  \int_{\vec{k}.}\sum_{i=0}^{\infty}-a_{h}(O_{k}^{i}O_{-k}^{i}+A_{k}%
^{i}A_{-k}^{i})+\label{f2first}\\
&  a_{h}\sum_{i,j}^{\infty}\left[  A_{k}^{i}A_{-k}^{j}K_{k}^{i,j}(1)+O_{k}%
^{i}A_{-k}^{j}L_{k}^{i,j}(1)+O_{-k}^{i}A_{k}^{j}M_{k}^{i,j}(1)+O_{k}^{i}%
O_{-k}^{j}N_{k}^{i,j}(1)\right]  \nonumber
\end{align}

We will use the degenerate perturbation theory similar to one used in Quantum
Mechanics to calculate the correction to the eigenvalues of matrix the LLL
states ( $A,O$ states) to order $a_{h}^{2}.$ The matrix $\widehat{H}%
\equiv\left(
\begin{array}
[c]{cc}%
K & L\\
M & N
\end{array}
\right)  $\bigskip is analogous to ''Hamiltonian'' while $\binom{A}{O}$ is
analogous to eigenvector. Explicitly the matrix elements are:%

\begin{align}
K_{k}^{i,j}(1)  &  =\frac{1}{\beta}<|\varphi|^{2}(\varphi_{-\overrightarrow
{k}}^{\ast i}\varphi_{-\overrightarrow{k}}^{j}+\varphi_{\overrightarrow{k}%
}^{i}\varphi_{\overrightarrow{k}}^{\ast j})-\frac{1}{2}(\varphi^{\ast2}%
\varphi_{\overrightarrow{k}}^{i}\varphi_{-\overrightarrow{k}}^{j}+\varphi
^{2}\varphi_{-\overrightarrow{k}}^{\ast i}\varphi_{\overrightarrow{k}}^{\ast
j})>\nonumber\\
N_{k}^{i,j}(1)  &  =\frac{1}{\beta}<|\varphi|^{2}(\varphi_{-\overrightarrow
{k}}^{\ast i}\varphi_{-\overrightarrow{k}}^{j}+\varphi_{\overrightarrow{k}%
}^{i}\varphi_{\overrightarrow{k}}^{\ast j})+\frac{1}{2}(\varphi^{\ast2}%
\varphi_{\overrightarrow{k}}^{i}\varphi_{-\overrightarrow{k}}^{j}+\varphi
^{2}\varphi_{-\overrightarrow{k}}^{\ast i}\varphi_{\overrightarrow{k}}^{\ast
j})>\label{kmatrix}\\
L_{k}^{i,j}(1)  &  =\frac{1}{\beta}<|\varphi|^{2}(\varphi_{-\overrightarrow
{k}}^{\ast i}\varphi_{-\overrightarrow{k}}^{j}-\varphi_{\overrightarrow{k}%
}^{i}\varphi_{\overrightarrow{k}}^{\ast j})+\frac{1}{2}(\varphi^{\ast2}%
\varphi_{\overrightarrow{k}}^{i}\varphi_{-\overrightarrow{k}}^{j}-\varphi
^{2}\varphi_{-\overrightarrow{k}}^{\ast i}\varphi_{\overrightarrow{k}}^{\ast
j})>\nonumber\\
M_{k}^{i,j}(1)  &  =\frac{1}{\beta}<|\varphi|^{2}(\varphi_{-\overrightarrow
{k}}^{\ast i}\varphi_{-\overrightarrow{k}}^{j}-\varphi_{\overrightarrow{k}%
}^{i}\varphi_{\overrightarrow{k}}^{\ast j})+\frac{1}{2}(\varphi^{\ast2}%
\varphi_{\overrightarrow{k}}^{i}\varphi_{-\overrightarrow{k}}^{j}-\varphi
^{2}\varphi_{-\overrightarrow{k}}^{\ast i}\varphi_{\overrightarrow{k}}^{\ast
j})>\nonumber
\end{align}

We will use definitions
\begin{align}
\beta_{k}^{n}  &  =<|\varphi|^{2}\varphi_{\overrightarrow{k}}\varphi
_{\overrightarrow{k}}^{\ast n}>\nonumber\\
\overline{\beta_{k}^{n}}  &  =<\varphi^{\ast}\varphi^{n}\varphi
_{\overrightarrow{k}}^{\ast}\varphi_{\overrightarrow{k}}>\label{beta}\\
\gamma_{k}^{n}  &  =<(\varphi^{\ast})^{2}\varphi_{-\overrightarrow{k}}%
\varphi_{\overrightarrow{k}}^{n}>\nonumber\\
\overline{\gamma_{k}^{n}}  &  =<\varphi^{\ast}\varphi^{\ast n}\varphi
_{\overrightarrow{k}}\varphi_{-\overrightarrow{k}}>\nonumber
\end{align}
When an index is zero we drop it throughout the paper. For example when $n=0,$
\ we call $\beta_{k}^{n}$ as $\beta_{k,}$ and when $k=0,\beta_{k}^{n}%
=\beta^{n}$ etc..).

Considering to order $a_{h}$ above matrix element $i,j=0$ of $K,L,M,N$ is:%

\begin{align}
K_{1}  &  =\frac{2}{\beta}\beta_{k}-\frac{1}{2\beta}\operatorname{Re}%
\gamma_{k}\nonumber\\
N_{1}  &  =\frac{2}{\beta}\beta_{k}+\frac{1}{2\beta}\operatorname{Re}%
\gamma_{k}\\
L_{1}  &  =M_{k}=-\frac{1}{\beta}\operatorname{Im}\gamma_{k}\nonumber
\end{align}
In deriving this we have used a property \bigskip$\beta_{k}=\beta_{-k}.$

We now diagonalize the matrix to find the eigenstates of LLL (which is of
order $a_{h}$). Eigenvalues are%

\begin{align}
\epsilon_{A}  &  =a_{h}\left(  -1+\frac{2}{\beta}\beta_{k}-\frac{1}{\beta
}|\gamma_{k}|\right) \label{elspec}\\
\epsilon_{O}  &  =a_{h}\left(  -1+\frac{2}{\beta}\beta_{k}+\frac{1}{\beta
}|\gamma_{k}|\right) \nonumber
\end{align}
as was found originally by Eilenberger \cite{Eilenberger}. The ''acoustic''
branch is shown on Fig.3a. The rotation transforming in to these eigenstates
is:
\begin{align}
\widetilde{A}_{k}  &  =\cos\frac{\theta_{k}}{2}A_{k}+\sin\frac{\theta_{k}}%
{2}O_{k}\label{rot}\\
\widetilde{O}_{k}  &  =-\sin\frac{\theta_{k}}{2}A_{k}+\cos\frac{\theta_{k}}%
{2}O_{k}\nonumber
\end{align}
where $\gamma_{k}\equiv|\gamma_{k}|\exp[i\theta_{k}]$. Similar calculation for
$n^{th}$ Landau level gives the spectrum:
\begin{equation}
\varepsilon_{A,O}^{n}=a_{h}\left(  -1+\frac{2}{\beta}<|\varphi|^{2}\varphi
_{k}^{\ast n}\varphi_{k}^{n}>\mp\frac{1}{\beta}|<(\varphi^{\ast})^{2}%
\varphi_{k}^{n}\varphi_{-k}^{n}>|\right)  \label{spec2}%
\end{equation}

\subsection{Spectrum of fluctuations beyond leading order in $a_{h}$}

In this subsection we calculate the correction of eigenvalues of LLL to order
$a_{h}^{2}$. The ''Hamiltonian'' $\widehat{H}$ in addition to has the $a_{h}$
part $\widehat{H_{1}}$ given in eq.(\ref{kmatrix}) also have the $a_{h}^{2}$
part $\widehat{H_{2}}$. As will be explained in the next section, we will need
only correction to the LLL to the $a_{h}^{2}$ order, not the HLL. Therefore we
will need only the $i,j=0$ matrix element of $\widehat{H_{2}}$:\bigskip%
\begin{align}
K_{2}  &  =\sum_{n=1}^{\infty}\frac{1}{nb\beta^{2}}\left[  \frac{3}{\beta
}\beta_{n}{}^{2}(2\beta_{k}-\operatorname{Re}\gamma_{k})-2\beta_{n}%
(\operatorname{Re}\overline{\beta_{k}^{n}}+\operatorname{Re}\overline
{\beta_{-k}^{n}}-\operatorname{Re}\overline{\gamma_{k}^{n}})\right]
\nonumber\\
N_{2}  &  =\sum_{n=1}^{\infty}\frac{1}{nb\beta^{2}}\left[  \frac{3}{\beta
}\beta_{n}^{2}(2\beta_{k}+\operatorname{Re}\gamma_{k})-2\beta_{n}%
(\operatorname{Re}\overline{\beta_{k}^{n}}+\operatorname{Re}\overline
{\beta_{-k}^{n}}+\operatorname{Re}\overline{\gamma_{k}^{n}})\right]
\label{ma1}\\
L_{2}+M_{2}  &  =\sum_{n=1}^{\infty}\frac{1}{nb\beta^{2}}\left[  -\frac
{6}{\beta}\beta_{n}^{2}\operatorname{Im}\gamma_{k}+\frac{4}{\beta}\beta
_{n}\operatorname{Im}\overline{\gamma_{k}^{n}}\right] \nonumber
\end{align}
Note that we do not show $L_{2}$ and $M_{2}$ separately as our result will
depend only on $L_{2}+M_{2}.$ According to the degenerate perturbation theory
we need to diagonalize $\widehat{H_{1}}$ which already has been done in the
previous subsection and then use the resulting states $\widetilde{A}_{k}$ and
$\widetilde{O}_{k}$ to calculate the second order correction to the
eigenvalue:$\varepsilon_{k}^{(2)}=a_{h}^{2}(E_{diag}+E_{offdiag})$. The
diagonal contribution is%

\begin{align}
E_{diag}  &  \equiv<\widetilde{A}_{k}|\widehat{H_{2}}|\widetilde{A}_{k}>\\
&  = (\cos\frac{\theta_{k}}{2})^{2}K_{2}+N_{2}(\sin\frac{\theta_{k}}{2}%
)^{2}+(L_{2}+M_{2})\sin\frac{\theta_{k}}{2}\cos\frac{\theta_{k}}{2}\nonumber
\end{align}

Substituting the matrix elements eq.(\ref{ma1})\ we obtain%

\begin{align}
E_{diag}  &  =\sum_{n=1}\frac{\beta_{n}}{nb\beta^{2}}{\LARGE \{}\frac
{3\beta_{n}}{\beta}(2\beta_{k}-|\gamma_{k}|)\nonumber\\
&  -2\left[  \operatorname{Re}\overline{\beta_{k}^{n}}+\operatorname{Re}%
\overline{\beta_{-k}^{n}}-\cos\theta_{k}\operatorname{Re}\overline{\gamma
_{k}^{n}}-\sin\theta_{k}\operatorname{Im}\overline{\gamma_{k}^{n}}\right]
{\LARGE \}} \label{epsilon2}%
\end{align}

\bigskip In the off diagonal contribution ,%

\begin{align}
E_{offdiag}  &  =-\sum_{n=1}\frac{<\widetilde{A}_{k}|\widehat{H_{1}%
}|n><n|\widehat{H_{1}}|\widetilde{A}_{k}>}{nb}\nonumber\\
&  =-\sum_{n}\frac{1}{nb}{\LARGE \{}(\cos\frac{\theta_{k}}{2})^{2}%
[|<A_{k}|\widehat{H_{1}}|A_{k}^{n}>|^{2}+|<A_{k}|\widehat{H_{1}}|O_{k}%
^{n}>|^{2}]\\
&  +(\sin\frac{\theta_{k}}{2})^{2}[|<O_{k}|\widehat{H_{1}}|O_{k}^{n}%
>|^{2}+|<O_{k}|\widehat{H_{1}}|A_{k}^{n}>|^{2}]+\sin\frac{\theta_{k}}{2}%
\cos\frac{\theta_{k}}{2}\nonumber\\
\times\lbrack &  <A_{k}|\widehat{H_{1}}|A_{k}^{n}><A_{k}^{n}|\widehat{H_{1}%
}|O_{k}>+<A_{k}|\widehat{H_{1}}|O_{k}^{n}><O_{k}^{n}|\widehat{H_{1}}%
|O_{k}>+c.c]{\LARGE \}}\nonumber
\end{align}

Details of calculation of these matrix elements can be found in Appendix B
together with definitions of quantities $F$. The result is:%

\begin{align}
E_{offdiag}  &  =-\frac{1}{b}\sum_{n}\frac{1}{n}\{\frac{1}{\beta^{2}}%
[|F_{k}^{n}(1)|^{2}+|F_{-k}^{n}(1)|^{2}+|F_{k}^{n}(2)|^{2}+|F_{-k}^{n}%
(2)|^{2}\nonumber\\
&  +\frac{\cos\theta_{k}}{\beta^{2}}[|F_{k}^{n}(1)|^{2}+|F_{-k}^{n}%
(1)|^{2}-|F_{k}^{n}(2)|^{2}-|F_{-k}^{n}(2)|^{2}]\label{epsilon3}\\
&  +\frac{2\sin\theta_{k}}{\beta^{2}}\operatorname{Im}[F_{-k}^{n}%
(1)F_{-k}^{\ast n}(2)-F_{k}^{\ast n}(1)F_{k}^{n}(2)]\}\nonumber
\end{align}
Similar result for $O_{k}$can be obtained from the above formula by changing
the sign of $\cos\theta_{k}$and the sign of $\sin\theta_{k}$in the formula above.

It is crucial to see whether there is an $k^{2}$ term in higher orders for the
''acoustic'' branch $A$. We calculated numerically the contributions to the
spectrum till $n=8$. All the $k^{2}$ contributions to any of them cancel.
Moreover even all the $k^{4}$ contributions for odd $n$ cancel although the
even $n$ give negative contribution to the rotationally symmetric combination
$(k_{x}^{2}+k_{y}^{2})^{2}$. Numerically the coefficients are: $2.2\cdot
10^{-6},5.0\cdot10^{-5},-6.3\cdot10^{-6},4.7\cdot10^{-7}$ for $n=2,4,6,8$
correspondingly. The resulting correction to the spectrum of ''acoustic''
branch due to the $n=2$ level is shown on Fig.3b.

After we have established the spectrum of elementary excitations of the
Abrikosov lattice, we are ready to calculate the fluctuations contributions to
various physical quantities.

\section{\bigskip Fluctuation contributions to free energy, magnetization and
specific heat}

\subsection{Higher Landau levels contributions to free energy}

The thermal fluctuation part is%

\begin{align}
-T\log[Z]  &  \equiv\mathcal{F}_{mf}+\mathcal{F}_{fluc};\nonumber\\
\mathcal{F}_{fluc}  &  =\frac{T}{2}\sum_{n=0}^{\infty}\left\{  Tr\ln\left[
\varepsilon_{A}^{n}(k)+\frac{k_{z}^{2}}{2}\right]  +Tr\ln\left[
\varepsilon_{O}^{n}(k)+\frac{k_{z}^{2}}{2}\right]  \right\} \label{e2}\\
&  =TL_{x}^{2}L_{z}\sigma b\sum_{n=0}^{\infty}\left[  <\sqrt{\varepsilon
_{A}^{n}(k)}>+<\sqrt{\varepsilon_{O}^{n}(k)}>\right] \nonumber
\end{align}
where $\sigma\equiv\frac{1}{\sqrt{2}2\pi}$ and we performed integration over
$k_{z}$.

The LLL contribution to order $\sqrt{a_{h}}$ in 2D has been calculated by
Eilenberger \cite{Eilenberger}. The 3D result is \cite{Rosenstein}:%

\begin{equation}
\frac{\mathcal{F}_{fluc}^{(1/2)}}{T}=\sigma ba_{h}^{1/2}\left[  <\sqrt
{\varepsilon_{A}^{(1)}(k)}>+<\sqrt{\varepsilon_{O}^{(1)}(k)}>\right]
=3.16\sigma ba_{h}^{1/2}%
\end{equation}
We calculated it's $\ $higher $a_{h}$ correction which is of order
$a_{h}^{3/2}$ using eq.(\ref{epsilon2}) and eq.(\ref{epsilon3})
\begin{equation}
\frac{\mathcal{F}_{fluc}^{(3/2)}}{T}=\frac{1}{2}\sigma ba_{h}^{3/2}\left[
<\frac{\varepsilon_{A}^{(2)}(k)}{\sqrt{\varepsilon_{A}^{(1)}(k)}}%
>+<\frac{\varepsilon_{O}^{(2)}(k)}{\sqrt{\varepsilon_{O}^{(1)}(k)}}>\right]
=-.445\sigma ba_{h}^{3/2}%
\end{equation}
As noted below $\varepsilon_{A}^{(2)}(k)$ and $\varepsilon_{A}^{(2)}(k)$ given
in the last section contain contributions from mixing with all the HLL. Table
1 details contributions to this term from levels till $n=8$. The contributions
are negative for all $n\neq6j$ where $j$ is an integer and positive otherwise.

\begin{center}
\textbf{Table 1.}

Contributions to free energy of mixing of the LLL with HLL.

Given in units of $\frac{1}{2}\sigma ba_{h}^{3/2}$.\bigskip%

\begin{tabular}
[c]{|c|c|c|c|c|c|c|c|c|}\hline
$\text{level }n$ & 1 & 2 & 3 & 4 & 5 & 6 & 7 & 8\\\hline
$A\text{ mode}$ & -.253 & -.082 & -.053 & -.063 & -.063 & .247 & -.017 &
-.005\\\hline
$O\text{ mode}$ & -.230 & -.086 & -.051 & -.023 & -.012 & .018 & -.003 &
-.001\\\hline
\end{tabular}
\bigskip
\end{center}

The contribution of HLL is:
\begin{align}
\frac{\mathcal{F}_{fluc}}{T}  &  =\sigma b\sum_{n=1}^{\infty}\left[
<\sqrt{nb+a_{h}\varepsilon_{A}^{n}(k)}>+<\sqrt{nb+a_{h}\varepsilon_{O}^{n}%
(k)}>\right]  \approx\nonumber\\
&  2\sigma b^{3/2}\sum_{n=1}^{\infty}\sqrt{n}+\frac{1}{2}\sigma b^{1/2}%
a_{h}\sum_{n=1}^{\infty}\frac{1}{\sqrt{n}}\left[  <\varepsilon_{A}%
^{n}(k)>+<\varepsilon_{O}^{n}(k)>\right]  +O(a_{h}^{2})\label{hllf}\\
&  =\frac{1}{T}(\mathcal{F}_{fluc}^{(0)}+\mathcal{F}_{fluc}^{(1)})\nonumber
\end{align}
The first divergent (as powers $3/2$ and $1/2$ in the ultraviolet) term
renormalizes energy. However it has a finite magnetic field dependant part
which should be calculated by subtracting the $b=0$ value of the free energy.
The proper regularization is made by restricting the number of Landau levels
and then showing that after regularization the answer does not depend on it.
The calculation is the same as one done in the normal phase (obviously UV
divergencies are insensitive to the phase in which they are calculated), see
for details and discussion \cite{Jana}. The result is:
\begin{equation}
\frac{\mathcal{F}_{fluc}^{(0)}}{T}=.526\sigma b^{3/2}%
\end{equation}
This exhibits the diamagnetic nature of the bosonic field \cite{Simon}. The
second term is proportional to $\sum\frac{1}{\sqrt{n}}$ and also diverges but
only as power 1/2 in the ultraviolet and renormalizes $a_{h}$. To see this we
calculate the sum
\begin{equation}
<\varepsilon_{A}^{n}(k)+\varepsilon_{O}^{n}(k)>_{k}=-2+\frac{4}{\beta
}<<|\varphi|^{2}\varphi_{k}^{\ast n}\varphi_{k}^{n}>_{x}>_{k}=-2+\frac
{4}{\beta}.\nonumber
\end{equation}
The last equality follows from the curious property of $\varphi_{k}^{\ast
n}(x)\varphi_{k}^{n}(x)$ that it depends only on $x_{i}\sqrt{b}-\varepsilon
_{ij}k_{j}$. We see that apart of renormalizations there is a finite
correction:
\begin{equation}
\frac{\mathcal{F}_{fluc}^{(1)}}{T}=1.459\left(  -1+\frac{2}{\beta}\right)
\sigma b^{1/2}a_{h}%
\end{equation}
The following one is of the order $a_{h}^{2}$ and will not be calculated here.

\section{Results for magnetization and specific heat. Range of applicability
of the loop expansion}

Here we discuss the nature and range of applicability of the expansions we
used for fluctuating superconductors (for which $Gi$ is not negligibly small).
There are two small parameters used. The \ first one is $\frac{a_{h}}{b}$
which controls the expansion of the mean field solution and therefore the HLL
corrections was already discussed in section IID. The second small parameter
controls the fluctuations. We assumed the mean field is the leading order and
then expanded the statistical sum around it. Summarizing all the corrections
the free energy density is
\begin{align}
\frac{\mathcal{F}}{T}  &  =\omega^{-1}a_{h}^{2}\left[  c_{2}^{(-1)}%
+c_{3}^{(-1)}\frac{a_{h}}{b}+c_{4}^{(-1)}\left(  \frac{a_{h}}{b}\right)
^{2}\right]  +\label{energyres}\\
&  \sigma ba_{h}^{1/2}\left[  c_{0}^{(0)}\left(  \frac{a_{h}}{b}\right)
^{-1/2}+c_{1/2}^{(0)}+c_{1}^{(0)}\left(  \frac{a_{h}}{b}\right)
^{1/2}+c_{3/2}^{(0)}\frac{a_{h}}{b}\right]  +\nonumber\\
&  \omega\sigma^{2}b^{2}a_{h}^{-1}\left[  c_{-1}^{(1)}\right]
\end{align}
where the coefficients (upper index is the power of $\omega$ and the lower
index is the power of $a_{h}$) are:%

\begin{align}
c_{2}^{(-1)}  &  =-0.434,c_{3}^{(-1)}=-0.0078,c_{0}^{(0)}=0.526,\\
c_{1/2}^{(0)}  &  =316,c_{1}^{(0)}=1.06,c_{3/2}^{(0)}=-0.445,c_{-1}%
^{(1)}=.118\nonumber
\end{align}
and $\omega$ is defined in eq.(\ref{omega}). The last term is the two loop
contribution calculated in I. One clearly see that $\omega ba_{h}^{-3/2}$
always appears together with an important ''loop factor'' $\sigma=\frac
{1}{2^{3/2}\pi}\simeq0.11$. The expansion parameter therefore is
\[
\sigma\omega ba_{h}^{-3/2}=\pi\sqrt{2Gi}\frac{tb}{\left(  1-t-b\right)
^{3/2}}=\frac{1}{\sqrt{2\pi}|a_{T}|^{3/2}}.
\]
In the last equation $a_{T}$ is the often used dimensionless LLL temperature
introduced by Thouless \cite{Thouless}. For $Gi=0.01$ the condition
$\sigma\omega a_{h}^{-3/2}<1$ is represented by the area above the dotted line
on Fig.1

Correspondingly the scaled magnetization is $m=-\frac{\partial\mathcal{F}%
}{\partial b}$:
\begin{align}
\frac{m}{T}  &  =\omega^{-1}\left(  c_{2}^{(-1)}a_{h}+\frac{3}{2}c_{3}%
^{(-1)}b^{-1}a_{h}^{2}+2c_{4}^{(-1)}b^{-2}a_{h}^{3}\right)  +\sigma
{\huge (}c_{3}^{(-1)}\omega^{-1}b^{-2}a_{h}^{3}-\frac{3}{2}c_{0}^{(0)}%
b^{1/2}-c_{1/2}^{(0)}a_{h}^{1/2}\nonumber\\
&  +\frac{1}{4}c_{1/2}^{(0)}ba_{h}^{-1/2}-\frac{1}{2}c_{1}^{(0)}b^{-1/2}%
a_{h}+\frac{1}{2}c_{1}^{(0)}b^{1/2}+\frac{3}{4}c_{3/2}^{(0)}a_{h}%
^{1/2}{\Huge )}-\omega\sigma^{2}\left(  2ba_{h}^{-1}+\frac{1}{2}b^{2}%
a_{h}^{-2}\right)  \left[  c_{-1}^{(1)}\right]
\end{align}
while the scaled specific heat is $c=-t\frac{\partial^{2}}{\partial t^{2}%
}\mathcal{F}$:%

\begin{align}
\frac{c}{t}  &  =tT_{c}\omega^{-1}\left(  -\frac{1}{2}c_{2}^{(-1)}\omega
^{-1}-3c_{3}^{(-1)}b^{-1}a_{h}-9c_{4}^{(-1)}b^{-2}a_{h}^{2}\right)  +\sigma
T_{c}{\huge (}\frac{1}{2}c_{1/2}^{(0)}ba_{h}^{-1/2}+\frac{1}{16}c_{1/2}%
^{(0)}bta_{h}^{-3/2}\nonumber\\
&  +c_{1}^{(0)}b^{1/2}+\frac{3}{2}c_{3/2}^{(0)}a_{h}^{1/2}-\frac{3}{16}%
c_{3/2}^{(0)}ta_{h}^{-1/2}{\Huge )}+T_{c}\omega\sigma^{2}b^{2}\left(
a_{h}^{-2}-\frac{t}{2}a_{h}^{-3}\right)  \left[  c_{-1}^{(1)}\right]
\end{align}

\section{Conclusion. \bigskip}

In this paper we showed why the LLL results are often valid far beyond the
naive limit of applicability of the approximation for both the mean field and
the fluctuation parts. Our results are valid strictly speaking between long
dashed line representing $H=\frac{H_{c2}(T)}{13}$ and one of the dashed curves
indicating the range of validity of the loop expansion for the fluctuation
contribution (depends on value of the Ginzburg number $Gi$). For
nonfluctuating strongly type II superconductors our results can be directly
checked by experiments done at low temperature or numerical solution (or even
the ''London limit approximation'') and are in clear agreement. For small, but
not very small Ginzburg parameter $Gi$ one can compare with existing Monte
Carlo simulations \cite{Sasik,Hu} or experiments. Of course one can use
existing high temperature expansion \cite{Thouless} to interpolate to the
present expansion range. Results for LLL were presented in I and the HLL do
not alter them significantly. The agreement with the MC simulations is very
good although obviously the melting transition is not seen. As argued in I it
is not expected to exist within the present model. The HLL do not change this
conclusion. On the contrary we explicitly showed that the ''supersoft'' $A$
mode has a propagator $1/(k_{z}^{2}+const(k_{x}^{4}+k_{y}^{4}))$ beyond the
LLL approximation laying to rest a suspicion that this is a fluke due to LLL.
This indicates that this unusual ''softness'' is due to some underlying
symmetry which have yet to be explicitly identified.

\section*{Acknowledgment}

We are grateful to our colleagues A. Knigavko, B. Bako, V. Yang. One of us
(B.R.) is grateful to G. Kotlar, A. Balatsky, L. Bulaevskii, Y. Kluger and R.
Sasik for numerous discussions in Los Alamos where this work started. The work
is part of the NCTS topical program on vortices in high Tc and was supported
by NSC of Taiwan.

\bigskip

\begin{center}
{\LARGE Appendix A: Third order correction to the mean field solution and free energy}
\end{center}

In this appendix we provide some details of the third order in $a_{h}$
calculation of the mean field solution of the GL equations.

To calculate $g_{2}^{n}$,one takes the inner product of $\varphi^{n}$on the
two sides of eq.(\ref{eq2}) and obtains eq.(\ref{g2n}). To calculate $g_{2}$,
we need to consider the GL equation to order $a_{h}^{7/2}$:
\begin{equation}
\mathcal{H}\Phi_{3}=\Phi_{2}-[(\Phi_{0})^{2}\Phi_{2}^{\ast}+(\Phi_{1})^{2}%
\Phi_{0}^{\ast}+2|\Phi_{0}|^{2}\Phi_{2}+2|\Phi_{1}|^{2}\Phi_{0}^{\ast}]
\end{equation}
Scalar product with $\varphi$ gives eq.(\ref{g2}).

Now we compute the $a_{h}^{4}$ order correction to free energy. Substituting
$\ \Phi_{2}$ from eqs.(\ref{g2n}) and (\ref{g2}) we find:%

\begin{align}
&  \frac{1}{2}a_{h}^{3}\left(  <\Phi_{2}|\mathcal{H}|\Phi_{1}>+<\Phi
_{1}|\mathcal{H}|\Phi_{2}>-<\Phi_{2}|\Phi_{0}>-<\Phi_{1}|\Phi2>-<\Phi_{1}%
|\Phi_{1}>\right) \label{free3}\\
&  =-\frac{g_{2}}{\beta^{1/2}}a_{h}^{3}+a_{h}^{3}\sum_{n=0}\left[  nbg_{1}%
^{n}g_{2}^{n}-\frac{1}{2}(g_{1}^{n})^{2}\right] \nonumber
\end{align}

Dominant contributions come from: $<6,6|0,0>=<0,0|6,6>=.80260$

$<6,0|6,0>=.80283$\bigskip and those coefficients are real.\bigskip\bigskip

\begin{center}
{\LARGE Appendix B: Second order correction to fluctuations spectrum}
\end{center}

In this appendix we list matrix elements of the correction $\widehat{H_{1}}$
given by eq.(\ref{kmatrix}) between various states used in the calculation of
the second order correction to energies of excitations.%

\begin{align}
<O_{k}|\widehat{H_{1}}|O_{k}^{n}>  &  =\frac{1}{\beta}<|\varphi|^{2}%
(\varphi_{-\overrightarrow{k}}^{\ast}\varphi_{-\overrightarrow{k}}^{n}%
+\varphi_{\overrightarrow{k}}\varphi_{\overrightarrow{k}}^{\ast n}%
)>\nonumber\\
&  +\frac{1}{2\beta}<\varphi^{\ast2}\varphi_{\overrightarrow{k}}%
\varphi_{-\overrightarrow{k}}^{n}+\varphi^{2}\varphi_{-\overrightarrow{k}%
}^{\ast}\varphi_{\overrightarrow{k}}^{\ast n}>\\
&  =\frac{1}{\beta}\left[  \beta_{-k}^{n}+\beta_{k}^{\ast n}+\frac{1}%
{2}\left(  \gamma_{-k}^{n}+\gamma_{k}^{\ast n}\right)  \right] \nonumber
\end{align}%

\begin{align}
<A_{k}|\widehat{H_{1}}|O_{k}^{n}>  &  =\frac{i}{\beta}<|\varphi|^{2}%
(\varphi_{\overrightarrow{k}}\varphi_{\overrightarrow{k}}^{\ast n}%
-\varphi_{-\overrightarrow{k}}^{n}\varphi_{-\overrightarrow{k}}^{\ast
})>+\nonumber\\
\frac{i}{2\beta}  &  <\varphi^{\ast2}\varphi_{\overrightarrow{k}}%
\varphi_{-\overrightarrow{k}}^{n}-\varphi^{2}\varphi_{-\overrightarrow{k}%
}^{\ast}\varphi_{\overrightarrow{k}}^{\ast n}>\\
&  =\frac{i}{\beta}\left[  -\beta_{-k}^{n}+\beta_{k}^{\ast n}+\frac{1}%
{2}\left(  \gamma_{-k}^{n}-\gamma_{k}^{\ast n}\right)  \right] \nonumber
\end{align}%

\begin{align}
<O_{k}|\widehat{H_{1}}|A_{k}^{n}>  &  =\frac{i}{\beta}\left[  \beta_{-k}%
^{n}-\beta_{k}^{\ast n}+\frac{1}{2}\left(  \gamma_{-k}^{n}-\gamma_{k}^{\ast
n}\right)  \right] \nonumber\\
<A_{k}^{n}|\widehat{H_{1}}|O_{k}>  &  =\frac{i}{\beta}\left[  \beta_{k}%
^{n}-\beta_{-k}^{\ast n}+\frac{1}{2}\left(  -\gamma_{-k}^{\ast n}+\gamma
_{k}^{n}\right)  \right] \label{elements}\\
<A_{k}|\widehat{H_{1}}|O_{k}^{n}>  &  =\frac{i}{\beta}\left[  \beta_{k}^{\ast
n}-\beta_{-k}^{n}+\frac{1}{2}\left(  -\gamma_{-k}^{n}+\gamma_{k}^{\ast
n}\right)  \right] \nonumber\\
<O_{k}^{n}|\widehat{H_{1}}|O_{k}>  &  =\frac{1}{\beta}\left[  \beta_{-k}^{\ast
n}+\beta_{k}^{n}+\frac{1}{2}\left(  \gamma_{-k}^{\ast n}+\gamma_{k}%
^{n}\right)  \right] \nonumber
\end{align}

etc. From those formula, we can show%

\begin{align}
|  &  <A_{k}|\widehat{H_{1}}|A_{k}^{n}>|^{2}+|<A_{k}|\widehat{H_{1}}|O_{k}%
^{n}>|^{2}=\frac{2}{\beta^{2}}\left[  |F_{k}^{n}(1)|^{2}+|F_{-k}^{n}%
(1)|^{2}\right] \nonumber\\
|  &  <O_{k}|\widehat{H_{1}}|O_{k}^{n}>|^{2}+|<O_{k}|\widehat{H_{1}}|A_{k}%
^{n}>|^{2}=\frac{2}{\beta^{2}}\left[  |F_{k}^{n}(2)|^{2}+|F_{-k}^{n}%
(2)|^{2}\right] \\
&  <A_{k}|\widehat{H_{1}}|A_{k}^{n}><A_{k}^{n}|\widehat{H_{1}}|O_{k}%
>+<A_{k}|\widehat{H_{1}}|O_{k}^{n}><O_{k}^{n}|\widehat{H_{1}}|O_{k}%
>+c.c=\nonumber\\
&  \frac{2}{\beta^{2}}[F_{k}^{\ast n}(1)F_{k}^{n}(2)-F_{-k}^{n}(1)F_{-k}^{\ast
n}(2)]+c.c,\nonumber
\end{align}

where%

\begin{align}
F_{k}^{n}(1)  &  =\beta_{k}^{n}-\frac{1}{2}\gamma_{k}^{n}\\
F_{k}^{n}(2)  &  =\beta_{k}^{n}+\frac{1}{2}\gamma_{k}^{n}\nonumber
\end{align}

Finally we can show that%

\begin{align}
E_{offdiag}  &  =-\frac{1}{b}\sum_{n}\frac{1}{n}{\LARGE \{}\frac{1}{\beta^{2}%
}\left[  |F_{k}^{n}(1)|^{2}+|F_{-k}^{n}(1)|^{2}+|F_{k}^{n}(2)|^{2}+|F_{-k}%
^{n}(2)|^{2}\right] \nonumber\\
&  +\frac{\cos\theta_{k}}{\beta^{2}}[|F_{k}^{n}(1)|^{2}+|F_{-k}^{n}%
(1)|^{2}-|F_{k}^{n}(2)|^{2}-|F_{-k}^{n}(2)|^{2}]\\
&  +\frac{\sin\theta_{k}}{\beta^{2}}2\operatorname{Im}[F_{-k}^{n}%
(1)F_{-k}^{\ast n}(2)-F_{k}^{\ast n}(1)F_{k}^{n}(2)]{\LARGE \}}\nonumber
\end{align}

\newpage

\begin{center}
{\Huge Figure captions}
\end{center}

\bigskip

{\LARGE Fig. 1}

\bigskip

The range of the validity of the expansions in $a_{h}$ and the loop expansion.
The region above the dotted line is the naively expected validity range of the
LLL approximation. The region above the long dashed line is the actual
validity range for the expansion of the mean field equations. The loop
expansion applicability range lies below the dashed curves. We plot two curves
with different values of Ginzburg number $Gi$, $Gi=0.1$ and $Gi=0.01$. The
validity combining the mean field expansion and the loop expansion lies
therefore between the long dashed line and the dashed curves.

{\LARGE Fig.2}

\bigskip

The density
$\vert$%
$\psi^{2}$%
$\vert$%
of the mean field Abrikosov solution for $b=0.1,t=0.5$. Fig.2a is the lowest
order approximation (LLL). Fig.2b is the solution with the next order
correction included, while in Fig.2c the next next order correction is included.

{\LARGE Fig.3}

The shear mode $A$ spectrum. Fig.3a is the spectrum obtained within the LLL
approximation. Fig.3b is the correction to the spectrum when the HLL mixing
effect is considered.
\end{document}